\documentclass[aps,a4paper,pre,showpacs,floats,nofootinbib,superscriptaddress,twocolumn]{revtex4}

\usepackage{amsmath}
\usepackage{amsfonts}
\usepackage{amssymb}
\usepackage[dvipdf]{graphicx}
\usepackage{subfigure}
\usepackage[pdfborder={0 0 0}]{hyperref}  
\usepackage{color}
\usepackage[latin1]{inputenc}
\usepackage{float}

\begin{document}
 \title{Quasi-stationary analysis  of the contact process on annealed
   scale-free networks}
 
 \begin{abstract}
   We present an analysis of the quasi-stationary (QS) state of the
   contact process (CP) on annealed scale-free networks using a
   mapping of the CP dynamics in a one-step processes and analyzing
   numerically and analytically the corresponding master equation.
   The relevant QS quantities determined via the master equation
   exhibit an excellent agreement with direct QS stochastic
   simulations of the CP.  The high accuracy of the resulting data
   allows to probe the strong corrections to scaling present in both
   the critical and supercritical regions, corrections that mask the
   correct finite size scaling obtained analytically by applying an
   exact heterogeneous mean-field approach.  {Our results
   represent a promising starting point for a deeper understanding
   of the contact process and absorbing phase transitions 
   on real (quenched) complex networks}
\end{abstract}
 
\author{Silvio C. Ferreira} \email{silviojr@ufv.br}\thanks{On leave at Departament de F\'{\i}sica i Enginyeria Nuclear, Universitat Polit\`ecnica de Catalunya, Barcelona, Spain.}

\author{Ronan S. Ferreira}

\affiliation{Departamento de F\'{\i}sica, Universidade Federal de
  Vi\c{c}osa, 36571-000, Vi\c{c}osa - MG, Brazil}

\author{Romualdo Pastor-Satorras} 
\affiliation{Departament de
  F\'{\i}sica i Enginyeria Nuclear, Universitat Polit\`ecnica de
  Catalunya, Campus Nord B4, 08034 Barcelona, Spain}

\pacs{89.75.Hc, 05.70.Jk, 05.10.Gg, 64.60.an}

\maketitle

\section{Introduction}
\label{sec:intro}

Complex network theory represents a general unifying formalism under
which is possible to understand and rationalize the intricate
connectivity and interaction patterns of many natural and man-made
systems. Thus, a systematic statistical analysis of large scale
dataset has allowed to unveil the existence of apparently universal
topological features, shared by a large number of different systems
from the technological, social or biological domains
\cite{barabasi02,mendesbook,newman2003saf}. Among these
characteristics, probably the most intriguing is the discovery of the
apparently ubiquitous scale-free (SF) nature of the connectivity
described by a probability $P(k)\sim k^{-\gamma}$ that an
element (vertex) is connected to other $k$ elements (has degree $k$),
with a degree exponent usually in the range $2 < \gamma < 3$
\cite{barabasi02,newman2003saf}. These and other discoveries have
promoted a large modeling activity, aimed at understanding the origin
and nature of the observed topological features
\cite{mendesbook,newman2003saf}. In recent years,
the research community has also devoted a great deal of attention to
the study of the dynamical processes on complex networks
\cite{barratbook,dorogovtsev07:_critic_phenom}, which can have
important implications in the understanding of real
processes such as the spread of epidemics in social systems
\cite{anderson92} or traffic in technological systems as the Internet
\cite{romuvespibook} or transport infrastructures \cite{Barrat:2004b}.

{The theoretical understanding of dynamical processes on complex
  networks is based in the application of mean-field approaches that
  are essentially based in the annealed network approximation
  \cite{dorogovtsev07:_critic_phenom}. Any non-weighted network is fully
  characterized by its adjacency matrix $a_{ij}$, taking the value $1$
  when vertices $i$ and $j$ are connected by an edge, and zero
  otherwise. In real (\textit{quenched}) networks, the values of the
  adjacency matrix are fixed and do not change with time.  When a
  dynamical process takes place on top of such a network, one is
  considering the network as frozen, with respect to the
  characteristic time scale $\tau_D$ of the dynamics. Other
  networks, however, are dynamical objects, changing in time over a
  characteristic time scale $\tau_N$. In this case, the adjacency
  matrix is defined only in a statistical sense, and a complete
  description of the network can be given in terms of its probability
  distribution $P(k)$ and its degree correlations $P(k|k')$
  \cite{marian1}.  In the limit $\tau_N \ll \tau_D$, when all edges
  are completely reshuffled between any two dynamical time steps, the
  resulting network is called \textit{annealed}
  \cite{boguna09:_langev}. Annealed networks represent an extremely
  important theoretical tool, because mean-field predictions for
  dynamical processes turn out to be exact in this kind of substrates
  \cite{dorogovtsev07:_critic_phenom}.}

Dynamical processes with absorbing configurations constitute a subject
of outstanding interest in non-equilibrium Statistical Physics
\cite{marro1999npt,Henkel} that have also found a place in network
science, as representative models of practical problems ranging from
epidemic spreading \cite{pv01a,PhysRevLett.105.218701},
infrastructure's resilience
\cite{PhysRevLett.85.5468,PhysRevLett.85.4626}, or activated dynamics
\cite{PhysRevLett.91.148701}, to mention just a few. The simplest
lattice model allowing absorbing configurations is the classical
contact process (CP) \cite{harris74}, whose universality class and
mean-field (MF) description have been discussed along the last few
years \cite{Castellano:2006, Hong:2007, Castellano:2008,
  PhysRevE.79.056115, boguna09:_langev}. In the CP defined in an
arbitrary network, vertices can be in two different states, either
empty or occupied. The dynamics includes the spontaneous annihilation
of occupied vertices, which become empty at unitary rate, and the
self-catalytic occupation of an empty vertex $i$ with rate $\lambda
n_i/k_i$, where $n_i$ is the number of occupied neighbors of $i$ and
$k_i$ is its degree. The model is thus characterized by a phase
transition at a value of the control parameter $\lambda = \lambda_c$,
separating an active from an absorbing phase devoid of active
vertices. Despite of its simplicity, the CP on SF networks exhibits a
very complex critical behavior even if studied in an annealed
substrate
\cite{Castellano:2008,PhysRevE.79.056115,boguna09:_langev}. This case
is particularly interesting, since in annealed networks all
connections are rewired at a rate much larger than the typical rates
involved in the dynamical process, implying that dynamical
correlations are absent \cite{boguna09:_langev}. In this case, the MF
approach is expected to be an exact description of the problem.

The configuration in which all vertices are empty plays a very
particular role, since once the system has fallen into this state, the
dynamics becomes frozen. For this reason, these states are called
absorbing and constitute a central feature in the analysis of finite
size systems since, in this case, the single actual stationary state is
the absorbing one \cite{marro1999npt}.  Finite size and absorbing
states must therefore be handled using suitable strategies,
concomitantly with an ansatz for the finite size scaling (FSS)~\cite{cardy88}. 
A widely adopted procedure is the so called
quasi-stationary (QS) state \cite{PhysRevE.71.016129,DickmanJPA}, in which the
absorbing configuration is suitably excluded from the dynamics.

In this work, we present a study of the QS state of the CP on SF annealed
networks, combining the QS numerical approach developed in
Ref.~\cite{PhysRevE.71.016129}, suitably extended to complex networks,
with the theoretical analysis of an approximated one-step process
derived from mean-field theory \cite{Castellano:2008}. Our analysis
allows us to determine the probability distribution of
activity both close to the critical point and in the off-critical
regime, as well as to obtain high quality data for relevant QS
quantities, such as the density of active sites or the characteristic
times. This last information is used to check the finite-size scaling
forms derived from mean-field theory, which turn out to be loaded with
very strong corrections to scaling.

{The results presented in our paper provide a deeper
  understanding of the nontrivial dynamics of the contact process in
  annealed networks. Moreover, they open the path to extension of the
  QS approach to the analysis of other dynamic processes with
  absorbing phase transitions taking place on more complex and/or
  realistic substrates such as, for example, quenched networks and
  small-world topologies \cite{watts98}, where edges are never
  rewired and dynamic correlations are usually present.}

We have organized our paper as follows. In Sec.~\ref{sec:QS} we review
the necessary background for QS analysis and numerical simulations,
while in Sec.~\ref{sec:CPannealed} we summarize the main results of MF
theory for the CP on annealed networks. A master equation approach for
the QS state is developed in Sec.~\ref{sec:QSCP}.  Section
\ref{sec:QScritic} is devoted to discuss the finite-size scaling forms
of the relevant QS quantities, as well as the corrections to scaling at
criticality. The off-critical analysis is discussed in
Sec.~\ref{sec:off}. Finally, our concluding remarks are presented in
Sec.~\ref{sec:conclu}.

\section{Finite size and the quasi-stationary state}
\label{sec:QS} 

In finite systems, the absorbing state is a fixed point
that can be visited even in the supercritical phase due to
stochastic fluctuations. Numerical simulations of finite systems are
particularly sensitive to absorbing states and therefore suitable
simulation strategies are required. The standard procedure consists in
restricting the averages to those runs that did not visit the
absorbing configuration \cite{marro1999npt}, the so-called surviving
averages. From a mathematical point of view, it is useful to define
the quasi-stationary (QS) state that consists of the ensemble of
states accessed by the original dynamical process at long times
restricted to those not trapped into an absorbing one
\cite{PhysRevE.71.016129}. The intensive quantities in a QS ensemble
must converge to the stationary ones in the thermodynamic
limit. Thus, in the active phase, the lifespan grows exponentially
fast with the system size and the QS state becomes identical to the
stationary one. In the sub-critical phase, on the other hand, the
activity in the QS state corresponds only to a few $[\mathcal{O}(1)]$
particles fluctuating above the absorbing state implying a density
that vanishes inversely proportional to the system size.

Formally, the QS state is related with the original one in
the limit $t\rightarrow\infty$ by
\begin{equation}
  \label{eq:qsdef}
  P(\sigma,t)=P_s(t)\bar{P}(\sigma),
\end{equation}
where $\bar{P}(\sigma)$ is the QS probability associated to the state
$\sigma$ and $P_s(t)$ is the survival probability, i.e., the
probability that the system is active up to time $t$.  For a one-step
process, the state of the system is completely determined by the
number of occupied vertices $n$. Letting $P_n(t)$ be the probability
that the system has $n$ particles at time $t$, the QS distribution
$\bar{P}_n$ is given by $P_n(t)=P_s(t)\bar{P}_n$, for which the
normalization $\sum_{n\ge 1} \bar{P}_n =1$ applies. The probability of
visiting the absorbing state is redistributed among the active
configurations, proportionally to $\bar{P}_n$, which constitutes the essence
of the QS state \cite{PhysRevE.71.016129}.  Knowledge of the QS
distribution allows to compute the standard quantities associated to
this state. For example, the probability to visit the vacuum in the CP
is given by $\dot{P}_0=P_1$, independently of the network substrate.
Thus, it is straightforward to show that the survival probability and
the pre-absorbing state are related by $dP_s/dt=-\bar{P}_1 P_s$
providing a characteristic time scale \cite{DickmanJPA}
\begin{equation}
  \label{eq:1}
  \tau=\frac{1}{\bar{P}_1}.
\end{equation}
Analogously, the density of active sites in the QS state is given by
\begin{equation}
  \label{eq:2}
  \bar{\rho} = \frac{1}{N} \sum_{n\geq1} n \bar{P}_n.
\end{equation}

The standard numerical procedure to simulate the QS regime based on
averages over survival runs has a limited accuracy due to the very
rare achievement of surviving configurations at very large times.
Thus, for instance, the stationary densities are determined as a
plateau at long times in the curve $\bar{\rho}(t)$ that is usually noisy
and short close to or below the criticality due to the limited number of
independent runs computationally accessible.  The previous
interpretation of the QS state provides an alternative simulation
strategy, in which every time the system visits the absorbing state,
this configuration is replaced by an active one randomly taken from
the history of the simulation \cite{PhysRevE.71.016129}. For this
task, a list with $M$ active configurations is stored and constantly
updated. An update consists in randomly choosing a configuration in
the list and replacing it by the present active configuration with a
probability $p_{r}$. After a relaxation time $t_r$, the QS
distributions are determined during an averaging time $t_a$.  The
improved QS method has been successfully applied to accurately
determine the universality class of several models with absorbing
configurations
\cite{PhysRevE.71.016129,PhysRevE.73.036131,PhysRevE.78.031133}.

\section{CP in annealed networks}
\label{sec:CPannealed} 

The network in which dynamics takes place is assumed to be
{annealed and therefore} completely characterized by the degree
distribution $P(k)$, the probability that a randomly chosen vertex has
a degree $k$, and the degree correlation function $P(k'|k)$ defined as
the conditional probability that a vertex of degree $k$ is connected
to a vertex of degree $k'$ \cite{marian1}. The number of vertices of
the network is denoted by $N$ and its maximum degree (cutoff) by $k_c$
\cite{mariancutofss}.  In an annealed framework, the MF rate equation
for the density of occupied vertices in the degree class $k$ (i.e. the
probability that a vertex of degree $k$ is occupied) is given by
\cite{Castellano:2008}
\begin{equation}
  \label{eq:rho_k}
  \frac{d}{dt}\rho_{k}(t)=-\rho_{k}(t)+\lambda
  k[1-\rho_{k}(t)]\sum_{k'}\frac{P(k'|k)\rho_{k'}}{k'}. 
\end{equation}
The first term represents the spontaneous annihilation and the second
one the creation inside the compartment $k$ due to the interaction
with all compartments under the hypothesis that there are no dynamical
correlations. A simple linear stability analysis allows to show the
presence of a phase transition, located at the value $\lambda_c=1$ and
independent of the degree distribution and degree correlations,
separating an active from an absorbing phase with $\rho_k=0$
\cite{boguna09:_langev}.  Considering in addition uncorrelated
networks with $P(k'|k) = k'P(k')/\langle k \rangle$ \cite{mendesbook},
the overall density $\rho = \sum_k \rho_k P(k)$ obeys the 
equation
\begin{equation}
  \label{eq:rhoglobal}
  \frac{d}{dt}\rho(t)=-\rho(t)+\lambda \rho\left[1- \langle k
    \rangle^{-1}\sum_{k}kP(k) \rho_{k}(t)\right] . 
\end{equation}

In Ref.~\cite{Castellano:2008}, it was realized that the low density regime of
Eq. (\ref{eq:rhoglobal}) can be understood
as a one-step process (biased random walk) with transition rates
\begin{equation}
\label{eq:ratesonestep}
 \begin{array}{lll}
  W(n-1,n) & = & n \\
  W(n+1,n) & = & \lambda n\left[1- \langle k \rangle^{-1}\sum_{k}kP(k)
    \rho_{k}(t)\right], 
 \end{array}
\end{equation}
where $W(m,n)$ corresponds to the transition from a state with $n$
particles to one with $m$ particles.  The stationary state $\partial_t
\rho_k = 0$ of Eq. (\ref{eq:rho_k}) reads as
\begin{equation}
  \label{eq:rhok}
  \bar{\rho}_k = \frac{\lambda k \bar{\rho}/\langle k
    \rangle}{1+\lambda k \bar{\rho}/\langle k \rangle}. 
\end{equation}
Close to the criticality, when the density at long times
is sufficiently small such that $\bar{\rho} k_c\ll 1$,
equation~(\ref{eq:rhok}) becomes $\bar{\rho}_k\simeq \lambda k\bar{\rho}/\langle
k \rangle$. Substituting in the transition rates, the first order
approximation for the one-step process is
\begin{equation}
\label{eq:rates_cp_dif}
 \begin{array}{lll}
  W(n-1,n) & = & n, \\
  W(n+1,n) & = & \lambda n(1-\lambda g n/N),
 \end{array}
\end{equation}
in which $g=\langle k ^2 \rangle/\langle k \rangle^2$. Based on
numerical evidences and scaling arguments, later confirmed by more
rigorous means in Refs. \cite{boguna09:_langev,PhysRevE.79.056115},
the authors proposed that the critical characteristic time $\tau$ and
stationary density $\bar{\rho}$ scale as \cite{Castellano:2008}
\begin{equation}
  \label{eq:taucrit}
  \tau\sim (N/g)^{1/2}
\end{equation} and 
\begin{equation}
  \label{eq:rhocrit}
  \bar{\rho}\sim (Ng)^{-1/2},
\end{equation} 
respectively.  For a network with degree exponent
$\gamma$ and a cutoff scaling with the system size as $k_c\sim
N^{1/\omega}$, where $\omega$ is an arbitrary positive parameter, the
factor $g$ scales for large $N$ as $g \sim k_c^{3-\gamma}$ for $2<\gamma<3$ 
and $g\sim\mbox{const}$ for $\gamma>3$. Therefore
the critical QS density scales as $\bar{\rho}\sim N^{-\hat{\nu}}$
where
\begin{equation}
  \label{eq:nu}
  \hat{\nu}  =
  \frac{1}{2}+\max\left(\frac{3-\gamma}{2\omega},0\right). 
\end{equation}
Similarly, the characteristic time follows $\tau\sim
N^{-\hat{\alpha}}$ with exponent 
\begin{equation}
  \hat{\alpha}=\frac{1}{2}-\max\left(\frac{3-\gamma}{2\omega},0\right) .
  \label{eq:5}
\end{equation}

The MF supercritical density for an infinite system was found to vanish at
criticality as $\bar{\rho}\sim \Delta^{\beta}$ where $\beta =
1/(\gamma-2)$ \cite{Castellano:2008} and $\Delta = \lambda-\lambda_c$.
For finite systems, the QS density has an anomalous cutoff-dependent FSS
given by \cite{boguna09:_langev}
\begin{equation}
  \bar{\rho}(\Delta,N) = \frac{1}{\sqrt{gN}}G\left(\Delta
    \sqrt{\frac{N}{g}}\right) \mbox{~~for~~} \frac{\Delta}{g} \ll  
  \frac{\lambda \langle k \rangle}{k_c},
 \label{eq:anomalousFSS}
\end{equation}
where $G(x) \sim x$ for $x\gg 1$ and $G(x)$ is constant for $x\ll
1$. The anomaly lies on the supercritical density dependence on the
system size through the factor $g$ given by 
$\bar{\rho}\sim \Delta/g$ if $\Delta>\sqrt{g/N}$ \cite{boguna09:_langev}.

\section{Master equation approach of the QS state}
\label{sec:QSCP}

In order to gain analytical information about the QS distribution not
far away from the critical point, we can consider the one-step process
approximation described by the transition rates in
Eq.~\eqref{eq:ratesonestep}. Starting from them, it is possible to
write down a master equation (ME) for the evolution of the number of particles
$P_n(t)$, taking the standard form
\begin{equation}
  \label{eq:3}
  \dot{P}_n = \sum_{m} W(n,m) P_m(t) - \sum_m W(m,n) P_n(t).
\end{equation}
In the long time limit we have $\dot{\rho}_k\approx 0$ and, consequently,
equation (\ref{eq:rhok}) can be applied resulting the ME
\begin{equation}
  \label{eq:master1}
  \dot{P}_n = (n+1)P_{n+1}+u_{n-1}P_{n-1}-(n+u_n)P_n,
\end{equation}
with $u_n =\lambda n (1-\Theta)$ and $\Theta$ given by 
\begin{equation}
  \label{eq:Theta}
  \Theta[\rho] = \frac{\lambda \rho}{\langle k \rangle^2} \sum_k
  \frac{k^2P(k)}{1+\lambda k \rho/\langle k \rangle}  .
\end{equation}
Substituting now $P_n(t)=P_s(t)\bar{P}_n$ and using
$dP_s/dt=-\bar{P}_1 P_s$ \cite{DickmanJPA}, the following recurrence
relation is obtained
\begin{equation}
  \label{eq:pniter}
  \bar{P}_{n} = \frac{1}{n}[(u_{n-1}+n-1-\bar{P}_{1})\bar{P}_{n-1} -
  u_{n-2}\bar{P}_{n-2}] ,
\end{equation} 
where $n=2,\cdots,N$ and $\bar{P}_0\equiv 0$. The QS distributions are
completely determined since $\bar{P}_1$, the initial condition to
iterate (\ref{eq:pniter}), is given by the normalization $\sum_{n\ge
  1}\bar{P}_n=1$.

Full information of the QS distribution can be obtained from the ME by
solving it numerically.  A numerical recipe to iterate the recurrence
relation is as follows \cite{DickmanJPA}: Start with a guess for
$\bar{P}^{(0)}_1$ and iterate (\ref{eq:pniter}) to find
$\bar{P}^{(0)}_n$, $n=2,\cdots,N$. Repeat the procedure using
$\bar{P}^{(j+1)}_1 = \bar{P}^{(j)}_1/\sum_n\bar{P}^{(j)}_n$ as new
guess until the normalization be reached. Suitable truncations can be
used to speed up the numerical process and to prevent
instabilities. The truncations at finite densities are justifiable
since the Central Limit Theorem guarantees that fluctuations much
larger or much smaller than the average are exponentially negligible.

In order to explore the properties of the QS state, we have performed
extensive Monte Carlo simulations of the CP on annealed networks. We
use a random neighbor network (RNN) \cite{boguna09:_langev} with
degree distribution $P(k)\sim k^{-\gamma}$, degree correlations
$P(k'|k) = k'P(k')/\langle k \rangle$, a degree cut-off $k_c =
N^{1/\omega}$, and a fixed minimum degree $k_0=2$. The single effect in
increasing the minimum degree $k_0$ is a shift to higher densities
which does not affect the critical properties.  In an annealed
approach, all links are redefined between any two time steps
in such a way that the neighbor of a given vertex is selected by
randomly choosing a vertex of the network with a probability
$k'P(k')/\langle k \rangle$. We perform stochastic simulations using
the usual scheme \cite{marro1999npt}: At each time step, an occupied
vertex is chosen at random and the time updated as $t\rightarrow
t+\Delta t$, where $\Delta t = 1/[(1+\lambda)n(t)]$ and $n(t)$ is the
number of occupied vertices at time $t$. With probability
$p=1/(1+\lambda)$, the occupied vertex becomes vacant. With
complementary probability $1-p=\lambda/(1+\lambda)$, one of its
neighbors (following RNN rules) is selected and, if empty,
occupied. If the selected neighbor is already occupied the simulation
goes to the next step. QS states were simulated using the method
described in Sec. \ref{sec:QS} with $M=10^3$, $p_{r}=0.02\Delta t$ and
$t_a=t_r=10^6$. Due the short distance between vertices, the
relaxation times are very short if compared with critical relaxation
on regular lattices.  The network sizes were varied from $10^3$ to
$10^7$ and $50-500$ network samples were used in the averages
(the larger the size of the network, the smaller the
number of samples). During the averaging interval,
the current configuration is counted in the QS distribution with a
probability proportional to its lifespan.

\begin{figure}[t]
  \centering
  \includegraphics[width=7.0cm,clip=true]{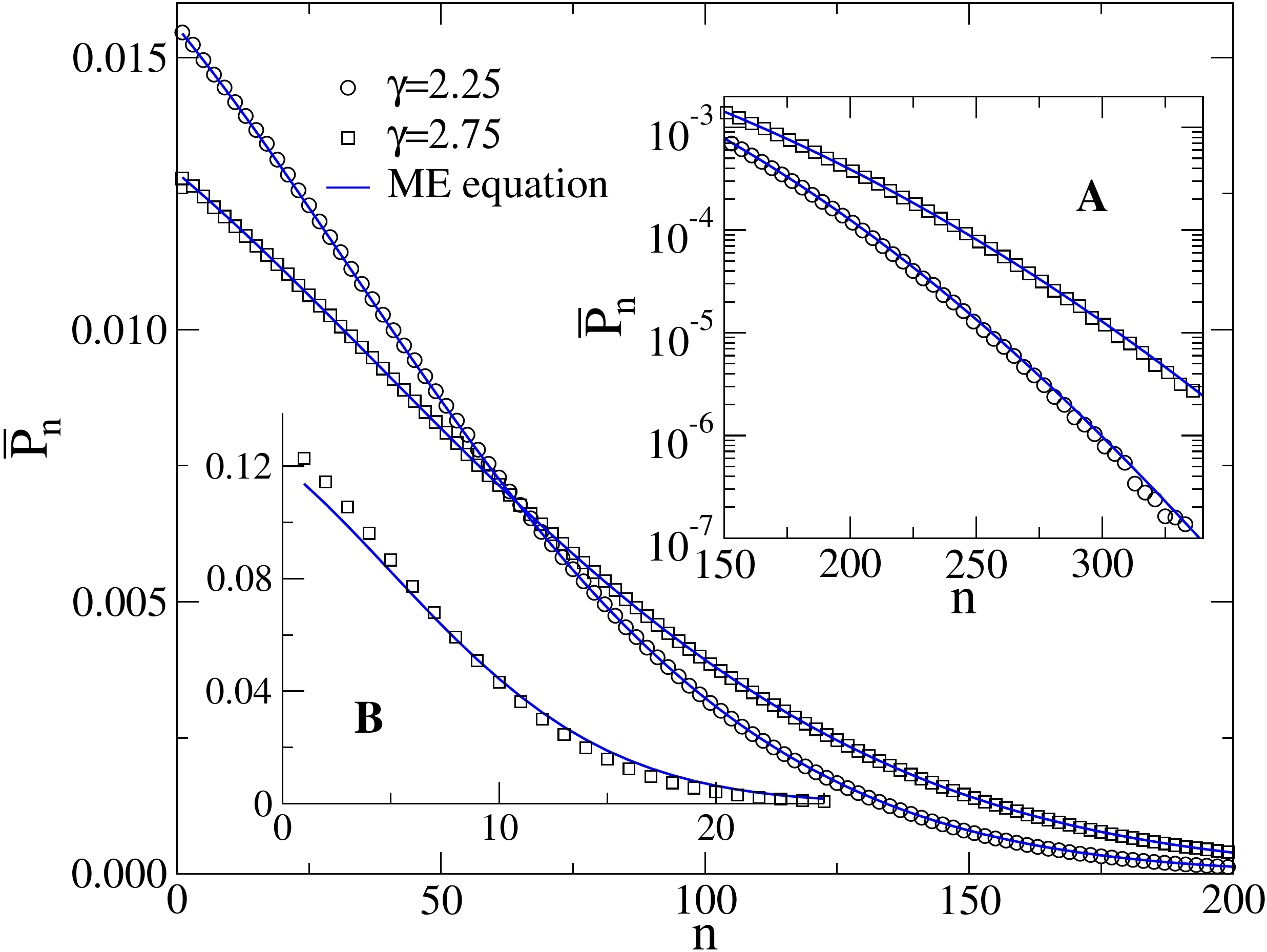}
  \caption{(Color on-line) QS probability distributions at criticality
    obtained in simulations of the CP on annealed SF networks with a
    cutoff $N^{1/2}$ are compared with the numerical solutions of
    Eq.~\eqref{eq:pniter}.  In the main plot the networks have size
    $N=2\times10^4$. Inset \textbf{A} shows the curves of the main
    plot in a logarithmic scale in order to compare the tails. Inset
    \textbf{B} shows the QS distributions for a smaller network of size
    $N=10^2$ with $\gamma=2.75$.}
  \label{fig:qs_vs_theor}
\end{figure}

The QS probability distributions for obtained in simulations of 
the critical CP are shown in
Fig. \ref{fig:qs_vs_theor} for different values of the degree exponent
$\gamma$ and network size $N$ and compared with the results of the
numerical solution of the recursion relation,
equation~\eqref{eq:pniter}.  A remarkable agreement between the simulations
and the one-step process approach is achieved even for the asymptotic
(Gaussian) tail as one can see in the inset \textbf{A} of
Fig. \ref{fig:qs_vs_theor}. A good accordance, which is improved as
size increases, is observed even for sizes as small as $10^3$, while
neat discrepancies appears for $N \sim 10^2$ (inset \textbf{B} of
Fig. \ref{fig:qs_vs_theor}). Actually, we can show that the one-step
and the Langevin approach developed in Ref. \cite{boguna09:_langev}
are equivalent in the low density limit (see Appendix) and,
consequently, the lower the density the better the one-step mapping.

\section{The QS state at criticality}

\label{sec:QScritic}

\subsection{Analytical approximation at criticality}
\label{sec:analyt-appr-at}

At criticality, where the densities at long times are very low, we have $u_n=\lambda
n (1-\lambda n/\Omega)$ where $\Omega = N/g$. In this limit,
equation (\ref{eq:master1}) corresponds exactly to the ME of the CP on a
complete graph of size $\Omega$, for which the QS analysis was already
worked out elsewhere \cite{DickmanJPA}. Analytical insights about the
criticality can be obtained through a van Kampen's expansion
\cite{vankampen} of the recurrence relation (\ref{eq:pniter}). Let us
consider the scaling solution of Eq.~\eqref{eq:pniter}
\begin{equation}
  \label{eq:qs_cpfull}
  \bar{P}_n=\frac{1}{\sqrt{\Omega}}f\left(\frac{n}{\sqrt{\Omega}}\right),
\end{equation}
where $f(x)$ is a scaling function to be determined.  Plugging
Eq.~\eqref{eq:qs_cpfull} into \eqref{eq:pniter}, and performing a
Taylor expansion up to second order, the result up to order
$\Omega^{-1}$ is
\begin{equation}
  \label{eq:eqdif_cpfull}
  x \frac{d^2 f}{dx^2}+(2+x^2)\frac{df}{dx}+2xf=-f_0f
\end{equation}
where $f_0 = \bar{P}_1\Omega^{1/2} =f(0)$ must be chosen to
impose the normalization condition $\int_0^\infty f(y) dy = 1$.  
Dickman and Vidgal \cite{DickmanJPA} analyzed
Eq. (\ref{eq:eqdif_cpfull}) numerically and checked the agreement with
the recurrence relation for the CP on the complete graph. We
complement the analysis by obtaining the asymptotic behaviors
analytically. It is straightforward to see that the distribution
decays linearly as $f(x) \simeq f_0 (1-f_0 x/2)$ for $x\ll 1$. A
correction to this initial behavior can be obtained discarding the
term $xf''(x)$ (a low curvature approximation) and the solution
satisfying the boundary condition $f(0)=f_0$ is
\begin{equation}
  \label{eq:xll1}
  f(x) \simeq \frac{2f_0}{2+x^2}\exp\left[-\frac{f_0}{\sqrt{2}}\arctan
    \left( \frac{x}{\sqrt{2}}\right)\right], \; x\ll1.
\end{equation}
For $x\gg 1$ the zeroth order terms are discarded and
Eq. (\ref{eq:eqdif_cpfull}) turns to $f''+xf'+2f \simeq 0$. The
solution satisfying the boundary condition $xf(x)\rightarrow 0$ for
$x\rightarrow\infty$ ($\langle x \rangle$ is finite) is 
\begin{equation}
  \label{eq:4}
  f(x)\sim \exp(-x^2/2) , \; x\gg1,
\end{equation}
implying a Gaussian tail.

\begin{figure}[t]
  \centering
  \includegraphics[width=7cm]{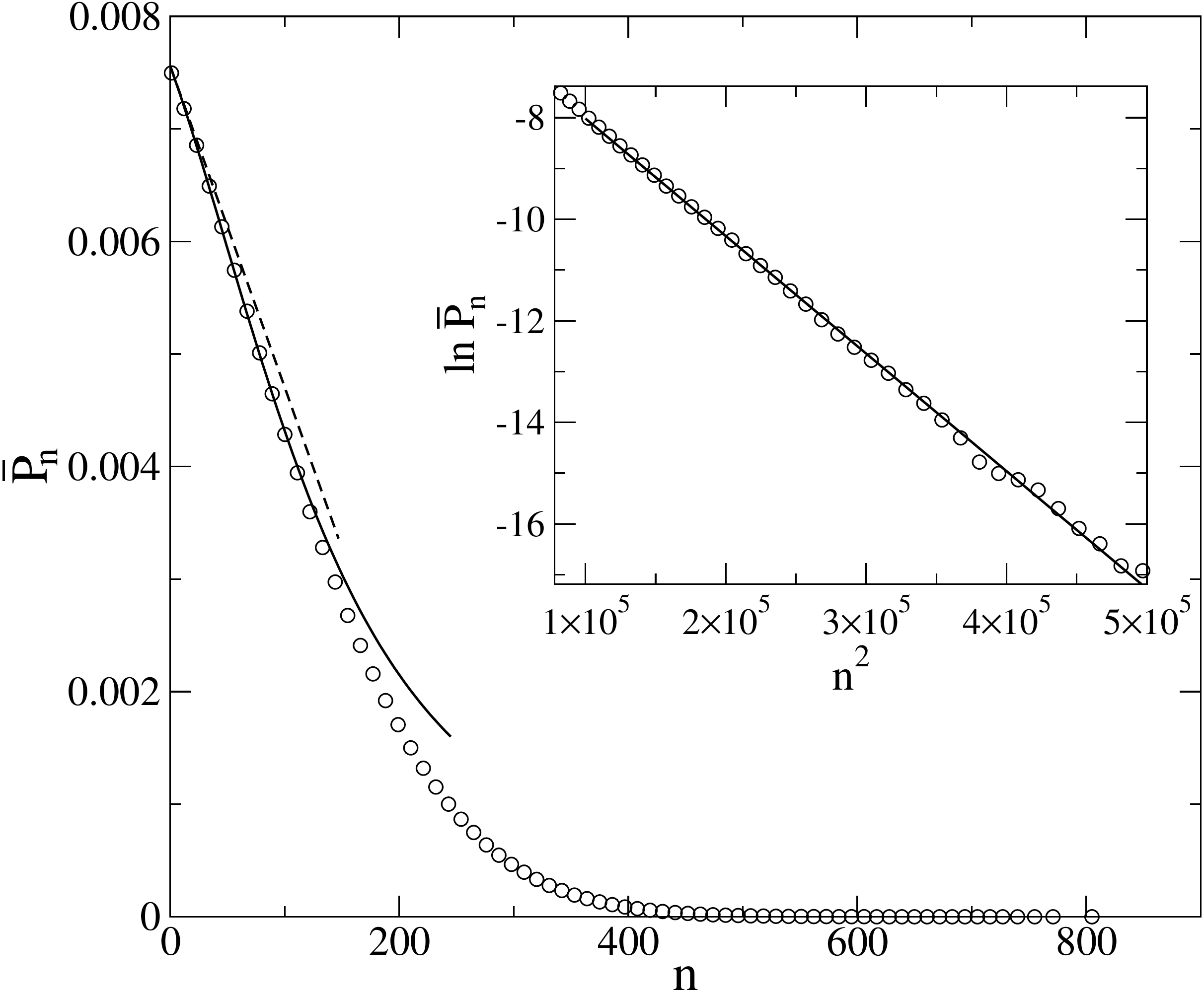}
  \caption{Comparison between the numerical simulations of the CP on
    annealed networks at criticality and the asymptotic behaviors of
    the QS distribution given by Eqs.~\eqref{eq:xll1} and
    \eqref{eq:4}. Numerical simulations for $N=160000$, $\gamma=2.25$
    and $\omega=2$ are represented by circles and the asymptotic
    solutions by lines.  The dashed line is the linear approximation
    for $n/\sqrt{\Omega}\ll 1$. The inset shows the comparison with
    the Gaussian tail using a straight line of slope $-1/2\Omega$.}
  \label{fig:compar}
\end{figure}

In Fig.~\ref{fig:compar} we compare the results of numerical
simulations of the CP at criticality ($\lambda=1$) on a network with
$N=160000$ nodes, degree exponent $\gamma=2.25$ and cutoff scaling exponent
$\omega=2$ with the corresponding analytical approximations.  Despite
of the lack of rigor, the analytical result is in good agreement with
numerical simulations.  The approximation (\ref{eq:xll1}) agrees with
numerical analysis even for $x \approx 1$ or equivalently for a number
of active vertices $n\approx\sqrt{\Omega}$. For the particular network of
Fig.~\ref{fig:compar} we have $\sqrt{\Omega}\approx 147$. The
accordance still holds for small networks ($\sim10^3$) independently
of the exponent degree and cutoff scaling.

The scaling function (\ref{eq:qs_cpfull}) encloses the FSS form of the
CP at criticality. In fact, the mean number of occupied vertices at
the QS regime is given by
\begin{equation}
  \bar{n} =
  \sum_{n=1}^{N}n\bar{P}_n=
  \sum_{n=1}^{N}\frac{n}{\sqrt{\Omega}}f\left(\frac{n}{\sqrt{\Omega}}\right),    
\end{equation}
where $\Omega=N/g$. Letting $x=n/\sqrt{\Omega}$ and $\Delta x =
1/\sqrt{\Omega}$, the sum can be approximated by a continuous
integration when $N\rightarrow\infty$, namely
\begin{equation}
  \label{eq:inte}
  \bar{n}= \Omega^{1/2} \sum_{x=m/\sqrt{\Omega}}^{N/\sqrt{\Omega}}
  xf(x) \Delta x \approx \Omega^{1/2}\int_0^\infty xf(x)dx \sim
  \Omega^{1/2}. 
\end{equation}
Therefore, the critical QS density is simply $\bar{\rho} \equiv
{\bar{n}}/{N} \sim(gN)^{-1/2}$, recovering the result firstly
presented in Ref.~\cite{Castellano:2008}. Analogously, the
characteristic time $\tau$ is also directly obtained by the present QS
analysis by noticing that, since $\bar{P}_1 = f_0 \Omega^{-1/2}$, we have
$\tau = 1/ \bar{P}_1 \sim\Omega^{1/2}=({N}/{g})^{1/2}$.

\subsection{Scaling at criticality}

The analysis of the QS state, either by direct QS simulations or
of the iterative solution of the corresponding approximate ME,
allows us to obtain high-quality data for the characteristic
quantities at criticality, namely the stationary density $\bar{\rho}$
and the characteristic time $\tau$. At criticality, these quantities
are expected to exhibit a scaling with system size of the form
$\bar{\rho}\sim N^{-\hat{\nu}}$ and $\tau\sim N^{\hat{\alpha}}$,
with exponents given by Eqs.~\eqref{eq:nu}
and~\eqref{eq:5} and depending on the cutoff scaling exponent $\omega$.

\begin{figure}[t]
  \centering
  \includegraphics[width=7cm]{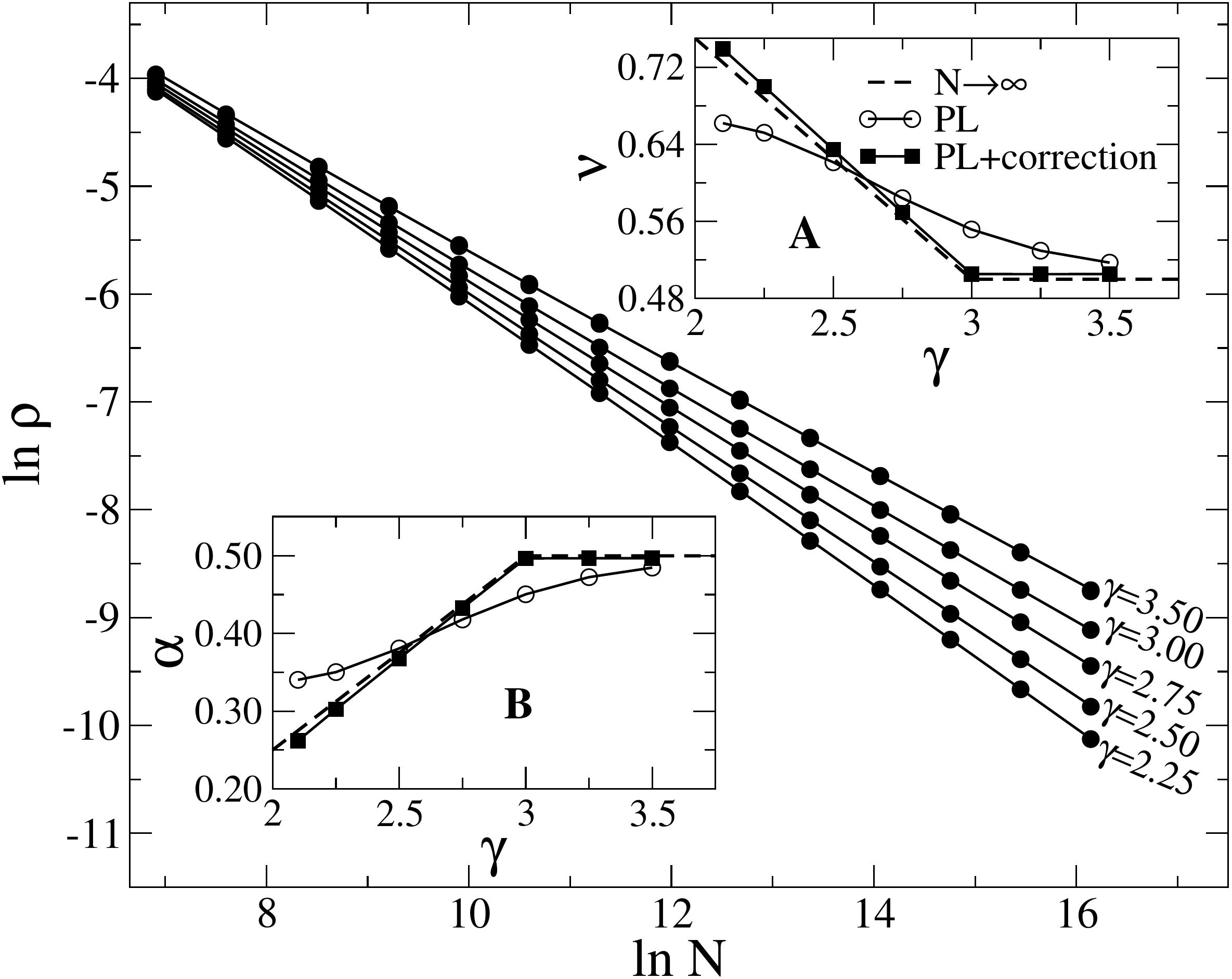}
  \caption{QS critical density for the CP on annealed SF networks with
    cutoff $k_c=N^{1/2}$. In main plot, the densities obtained with
    the numerical solution of the ME (lines) and QS simulations
    (symbols) are shown. Inset \textbf{A} shows the exponent of the
    scaling law $\bar{\rho}\sim N^{-\hat{\nu}}$ obtained analytically
    for asymptotically large systems ($N\rightarrow\infty$),
    performing a direct fit to a power-law form (PL) of the data in
    main figure, and performing a fit to a power with corrections to
    scaling.  Inset \textbf{B} shows the same analysis for the
    characteristic time $\tau\sim N^{\hat{\alpha}}$.}
 \label{fig:qs_critc}
\end{figure}

\begin{figure}[t]
  \centering
  \includegraphics[width=7cm]{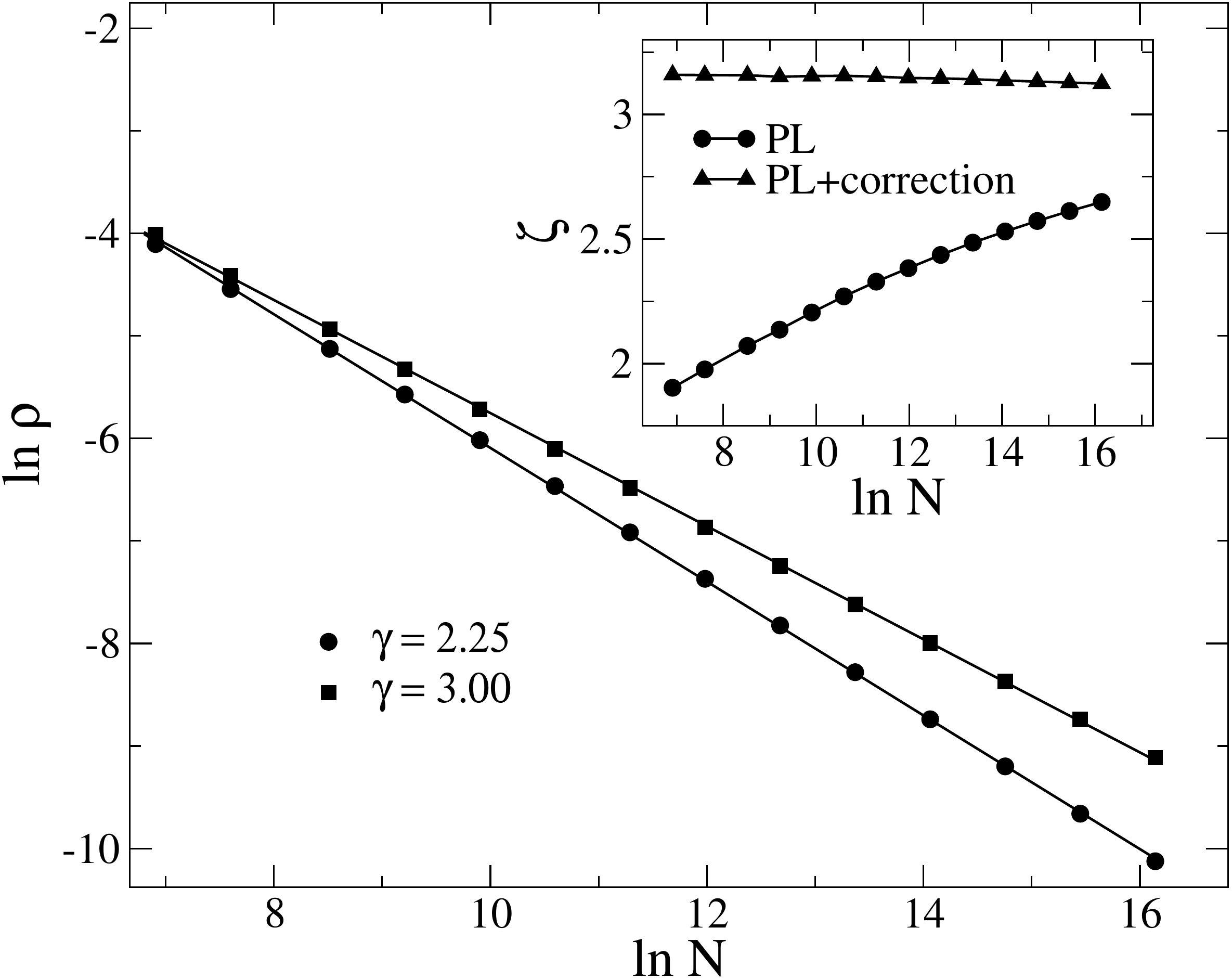}
  \caption{Power law regressions (solid lines) of the critical QS
    densities (symbols) for CP on annealed SF networks with cutoff
    $k_c=N^{1/2}$ and degree exponents $\gamma=2.25$ and
    $\gamma=3.00$. Inset shows the QS densities for $\gamma=2.25$
    rescaled by pure power law (PL), $\zeta=\bar{\rho} N^{0.6875}$,
    and a PL with correction to scaling, $\zeta=\bar{\rho}
    N^{0.6875}(1+2\times 2^{0.25} N^{-0.125}+3 \times 2^{0.5}
    N^{-0.25})^{0.5}$.}
 \label{fig:pl_correct}
\end{figure}

QS critical densities as functions of the
network size $N$, computed for several degree exponents and $\omega=2$
are shown in the main plot of Fig.~\ref{fig:qs_critc}. 
Again, an incontestable agreement between QS simulations and the 
numerical ME approach is observed. Similar agreement is obtained for the
analysis of characteristic time (data not shown). However, if
one tries to recover the theoretical scaling exponents, taking the
values $\hat{\nu}=\max[1/2,(5-\gamma)/4]$ and $\hat{\alpha} =
\min[1/2,(\gamma-1)/4]$, by means of a direct power law regression (insets
in Fig.~\ref{fig:qs_critc}), a very poor
agreement with the expected analytical exponents is observed, as already noted
in Ref.~\cite{PhysRevLett.98.029802}. Indeed, if we look
carefully at the data we can observe that,
even though a pretty good linear fit can be resolved for data range
corresponding to $10^3\le N \le 10^7$ as shown in Fig. \ref{fig:pl_correct},
the actual regressions are a little bit curved. 
Indeed, a careful analysis of the numerical data can resolve a slight
downward (negative) curvature for $\gamma\le 5/2$ and an slight upward
(positive) curvature for $\gamma>5/2$ at $\lambda=\lambda_c$. In a QS
analysis these behaviors usually indicate a system slightly out of the
critical point, being sub and supercritical for down and upward
curvatures, respectively. But this is not the case for the data shown
in Figs. \ref{fig:qs_critc} and \ref{fig:pl_correct}, since the MF
result is exact for the critical CP on annealed networks. Similar
behaviors occur for plots of $\tau$ versus $N$.

As noted in \cite{Castellano:2008,boguna09:_langev}, the origin of
this poor agreement between theory and simulations lies in the
implicit dependence on $N$ of the $g$ factor defining the size scaling
of $\bar{\rho}$ and $\tau$. In fact, the scaling forms $\bar{\rho}\sim
N^{-\hat{\nu}}$ and $\tau\sim N^{\hat{\alpha}}$ make only sense in
the limit of very large $N$, when $g$ has achieved its truly
asymptotic form. For intermediate values of $N$, instead, one should
keep the scaling forms with the simultaneous dependence on $g$ and $N$
\cite{Castellano:2008}. If we want instead to make explicit the
scaling with network size, we must consider that $g=\langle k^2\rangle
/ \langle k\rangle^2$ behaves, in the continuous degree limit, as
\begin{equation}
  \label{eq:gfull}
  g=\frac{(\gamma-2)^2 k_0^{\gamma-1}
  }{(\gamma-1)(3-\gamma)}
  \frac{(1-\xi^{\gamma-1})(1-\xi^{3-\gamma})}{(1-\xi^{\gamma-2})^2}k_c^{3-\gamma}, 
\end{equation}
where $\xi=k_0/k_c<1$. So, when $N\to\infty$,
$g\sim N ^{(3-\gamma)/\omega}$ for $2<\gamma<3$ and $g\sim$~const for
$\gamma>3$.

\begin{figure}[t]
  \centering
  \includegraphics[width=7cm]{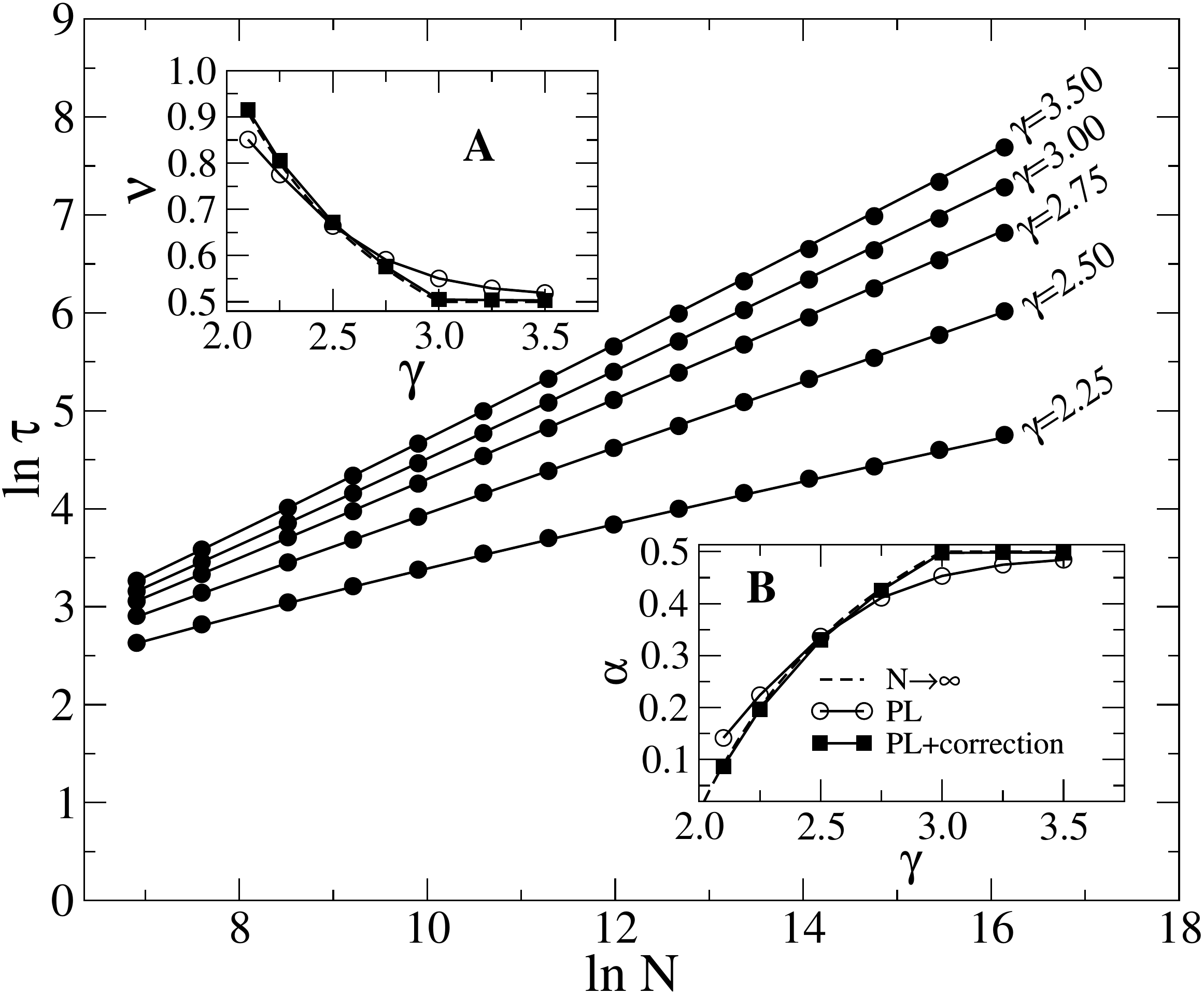}
  \caption{Critical QS characteristic times for CP on annealed SF
    networks with cutoff $k_c=N^{1/(\gamma-1)}$. In main plot, lines
    represent numerical solution of the ME and symbols QS simulations.
    The inset \textbf{A} shows the exponent of the scaling law 
    $\bar{\rho}\sim N^{-\hat{\nu}}$ while the inset \textbf{B} shows the 
    exponent of the scaling law $\tau\sim N^{\hat{\alpha}}$. Legends as in
    Fig. \ref{fig:qs_critc}}
  \label{fig:tauqs_nat}
\end{figure}

From Eq.~\eqref{eq:gfull} it is possible to work out the explicit form
of the corrections to scaling in a direct analysis of the QS quantities
as functions of the network
size. So, considering $2<\gamma<3$ and performing an expansion to
leading order in $\xi$, equation (\ref{eq:gfull}) yields
\begin{equation}
  \label{eq:gexp}
  g \simeq \mathrm{const} \times  \left(1 - \xi^{3-\gamma} +
    2\xi^{\gamma-2} \cdots\right) k_c^{3-\gamma}. 
\end{equation}
Substituting into the MF scaling result $\bar{\rho}\sim (gN)^{-1/2}$,
we obtain the expression for the stationary density
\begin{equation}
  \label{eq:correct}
  \ln \rho = C -\hat{\nu}\ln
  N+\frac{1}{2}\frac{k_0^{3-\gamma}}{N^{\frac{3-\gamma}{\omega}}} -
  \frac{k_0^{\gamma-2}}{N^{\frac{\gamma-2}{\omega}}}. 
\end{equation}
Notice that the corrections do not introduce any parameters to be
fitted. Similar expressions are found for $\gamma\ge
3$. Equation (\ref{eq:correct}) explains the deviations from the of
power law regime $\bar{\rho} \propto N^{-\hat{\nu}}$ observed for the
CP on annealed SF networks. It is easy to see that the leading term
for $2<\gamma\le 5/2$ is negative and causes a downward curvature, in
the same way that the leading term for $ \gamma> 5/2$ bends the curve
upwardly.  Even though the correction vanishes for
$N\rightarrow\infty$, it may occur extremely slowly due to the small
exponents involved. For $\gamma\approx 3$ and $\gamma\approx 2$ the
corrections are logarithmic and are thus relevant for any finite size.

Introducing the corrections given in Eq. (\ref{eq:correct}) in the
form
\begin{equation}
  \label{eq:6}
  \ln \rho' = \ln \rho
  -\left(\frac{1}{2}\frac{k_0^{3-\gamma}}{N^{\frac{3-\gamma}{\omega}}}
    - 
    \frac{k_0^{\gamma-2}}{N^{\frac{\gamma-2}{\omega}}}\right) = C -\hat{\nu}\ln
  N
\end{equation}
and performing a linear fit, the asymptotic exponent $\hat{\nu}$
is recovered as one can see in the inset \textbf{A} in
Fig. \ref{fig:qs_critc}.  Equivalent corrections can be easily
obtained for $\ln \tau$ \textit{vs.} $\ln N$ and the expected exponent
$\hat{\alpha}$ is recovered as shown in the inset \textbf{B} in
Fig. \ref{fig:qs_critc}.  Additional proof of the strong finite size
corrections is provided in the inset of Fig. \ref{fig:pl_correct}, in
which the critical QS density is rescaled by the predicted power law
(PL) with exponent $\bar{\nu}$ and by this same PL with the correction
(\ref{eq:correct}). The first case is clearly size dependent while
the second is flat. It is worth to note that the corrections are so
strong for $\gamma=2.25$ and $\omega=2$ that keeping only the leading
term $\mathcal{O}(N^{-0.125})$ was not enough to account for the
deviation.

We additionally performed the analysis for $\omega=\gamma-1$, which
corresponds to the natural cutoff that emerges in the absence of a
structural cutoff \cite{dorogovtsev07:_critic_phenom}. For sake of
simplicity, a hard cutoff was adopted such that connectivities larger
than $k_c=N^{1/(\gamma-1)}$ are forbidden.  For this cutoff, the
scaling exponents for the critical density and characteristic time are
$\hat{\nu}=\max[1/2,1/(\gamma-1)]$ and
$\hat{\alpha}=\max[1/2,(\gamma-2)/(\gamma-1)]$, respectively. Exactly
as in the $\omega=2$ case, the QS analysis via ME agrees with
simulations and the correct scaling exponents are obtained if
corrections to the scaling are considered.  Figure \ref{fig:tauqs_nat}
shows the characteristic time and the insets therein the scaling
exponent analysis for $\omega=\gamma-1$. Comparing the insets in
Figs.~\ref{fig:qs_critc} and \ref{fig:tauqs_nat}, we can observe that
the relative deviation between the exponents obtained using a simple
and a corrected PL is smaller for the natural ($\omega=\gamma-1$) than
for the cutoff $k_c=N^{1/2}$. Indeed, equation~(\ref{eq:correct}) tells
that the larger the cutoff exponent $\omega$, the stronger the
corrections to the scaling. However, even if the cutoffs are not
imposed, the natural one emerges spontaneously in networks with
power-law degree distributions
\cite{dorogovtsev07:_critic_phenom}. Consequently, these corrections
to the scaling may be also present in the CP and other dynamical
processes in SF substrates including the quenched case.

The previous results may have a remarkable impact in the analysis of
absorbing phase transitions in complex networks. The usual QS analysis
assumes a power law dependence of the order parameters with the size at
criticality. Such assumption is commonly used as a criterion to
determine the critical point of absorbing phase transitions in regular
lattices \cite{marro1999npt} and has been extended to quenched complex
networks \cite{Castellano:2006,Hong:2007}. Corrections to scaling in the
form $1-\mbox{const}\times N^{-0.75}$ were already observed in QS
simulations of the directed percolation universality class in
hypercubic lattices, including the contact process
\cite{PhysRevE.71.016129,Sander2009}. Since these corrections decay
with a large exponent, they are significant only for small systems. In
complex networks, the scenario is quite different since the
corrections, which emerge from the intrinsic SF nature of the substrate,
vanish very slowly and are important even for large systems ($N \sim
10^7$ in the present work).

\section{Off-critical QS analysis}
\label{sec:off}

\begin{figure}[t]
  \centering
  \includegraphics[width=7cm]{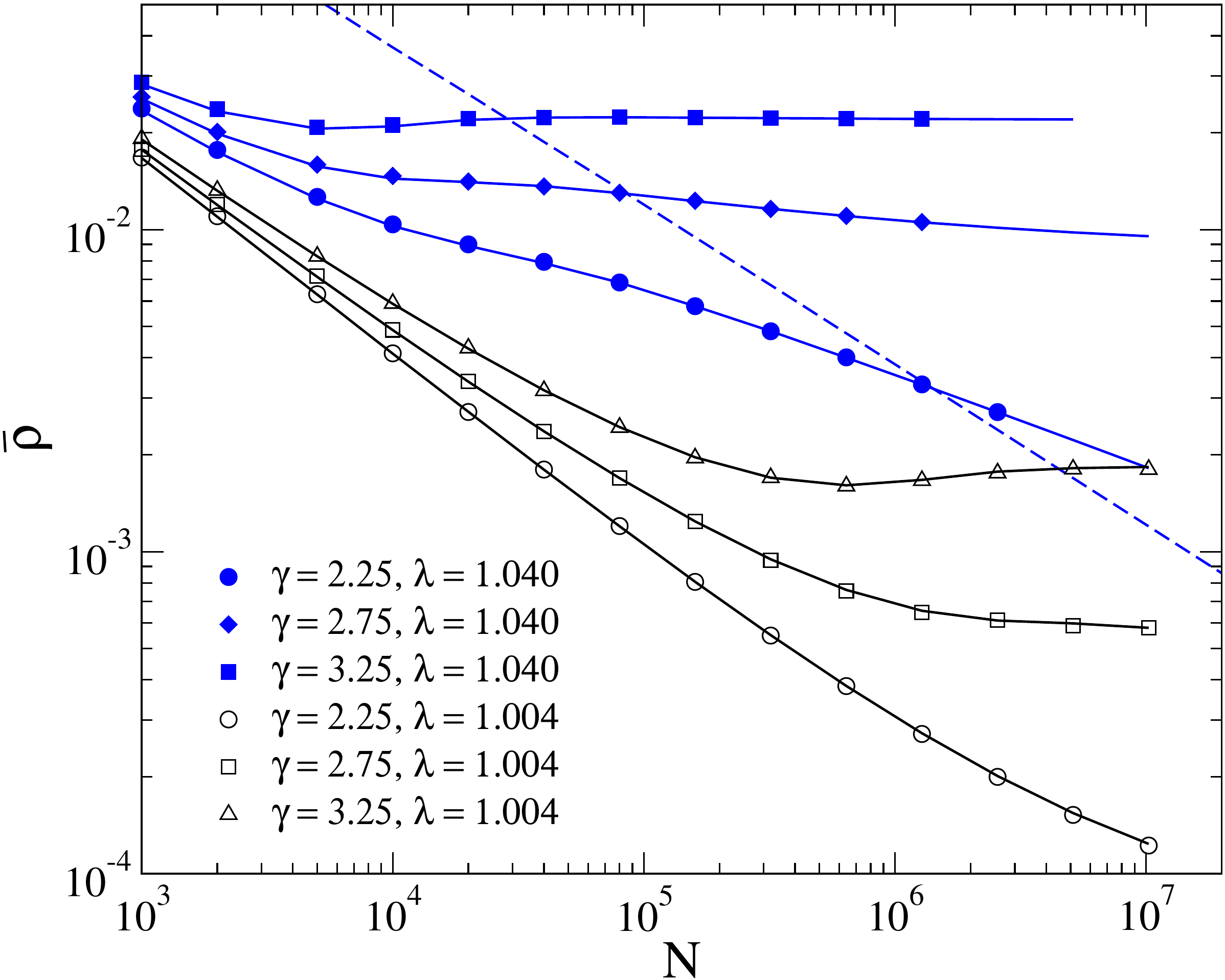}
  \caption{(Color on-line) Supercritical densities as functions of
    the system size for three degree distributions and a cutoff
    $k_c=N^{1/2}$. Solid lines are the ME numerical solutions and
    symbols QS simulations.} 
   \label{fig:qs_p04900}
\end{figure}

\begin{figure}[t]
  \centering
  \includegraphics[width=7cm]{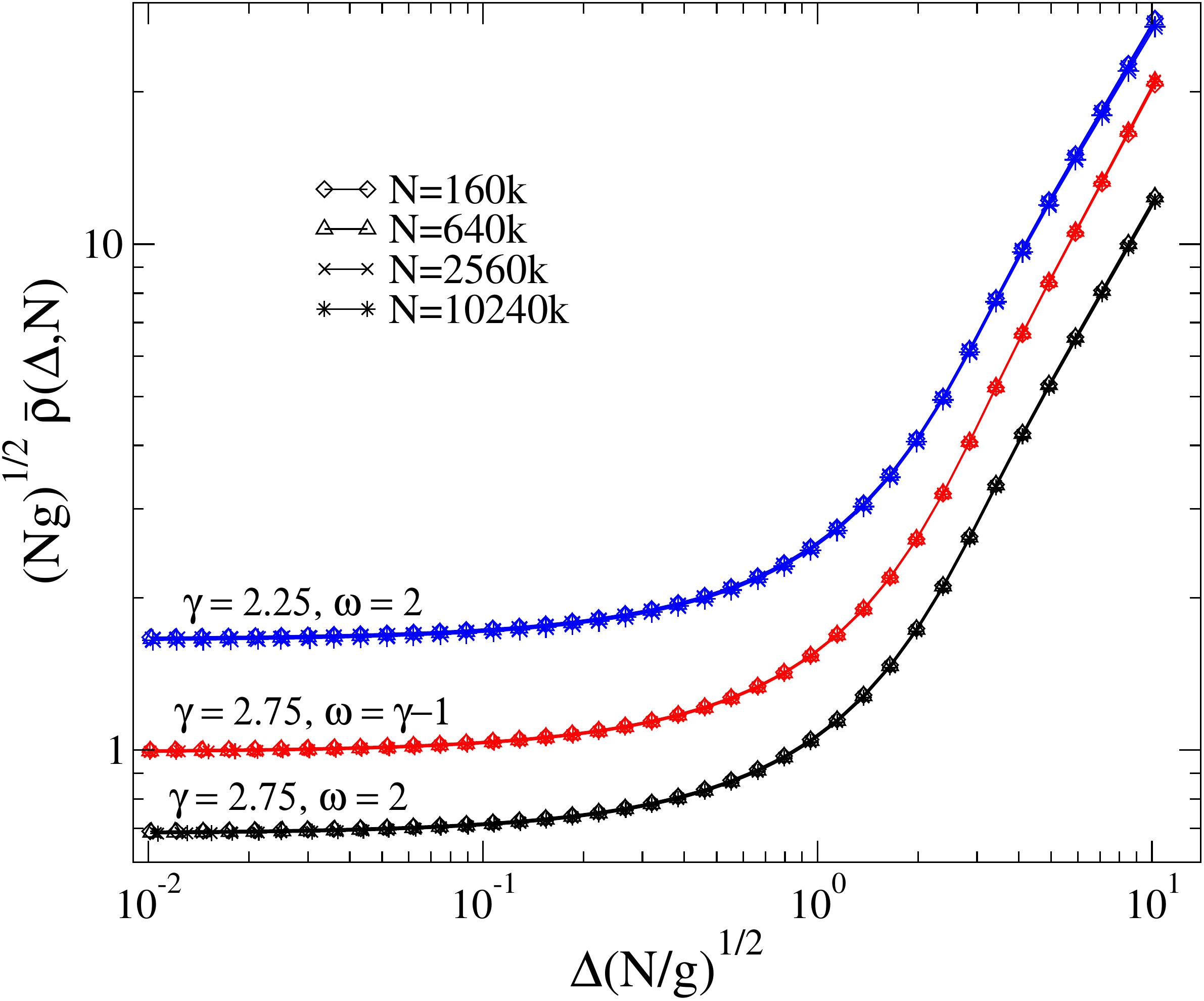}
  \caption{(Color on-line) Collapses of the numerical ME solution
    using the anomalous scaling function
    Eq.~(\ref{eq:anomalousFSS}). Densities obtained for network sizes
    $N=16\times10^4$, $64\times10^4$, $256\times10^4$, and
    $1024\times10^4$ are shown.  Data were shifted to avoid overlaps.
  }
 \label{fig:rholamb}
\end{figure}

The analysis performed for the critical CP in the previous sections is
expected to work also for the off-critical phase if the densities are
still sufficiently small. In Fig. \ref{fig:qs_p04900} we compare the
QS density obtained by ME iterative solution and numerical computed
for networks with different degree exponents. In the plot we explore
the supercritical regime with rates $\lambda=1.004$ and 1.040 (0.4\%
and 4\% above the critical point, respectively). The one-step ME
predicts QS densities very accurately even for a substantial distance
from the critical point, corroborating that the approach is also
suitable for the supercritical phase.  The anomalous FSS form in the
supercritical regime, equation~(\ref{eq:anomalousFSS}), depending on
$\Delta$, $N$ and $g$ simultaneously, is checked in
Fig. \ref{fig:rholamb}, where we present the collapses of the data 
obtained from iterative solutions of the ME. 
 Degree exponents $\gamma=2.25$ and $2.75$ using structural
($\omega=2$) and natural ($\omega=\gamma-1$) cutoffs are shown. Excellent
collapses are obtained in all cases, in agreement with
Refs. \cite{boguna09:_langev,PhysRevE.79.056115}.

\begin{figure}[t]
  \centering
  \includegraphics[width=7cm]{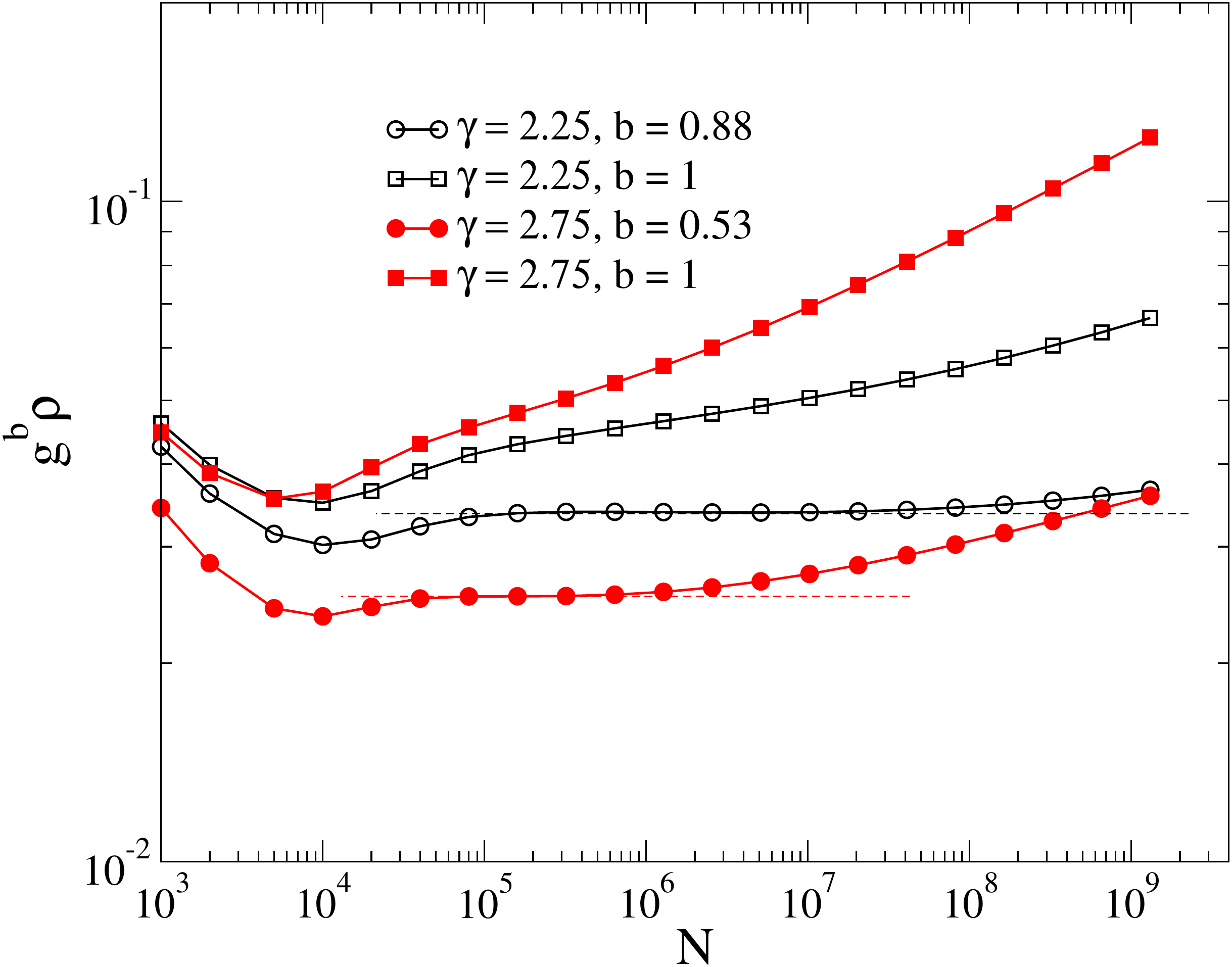}\\
  \includegraphics[width=7cm]{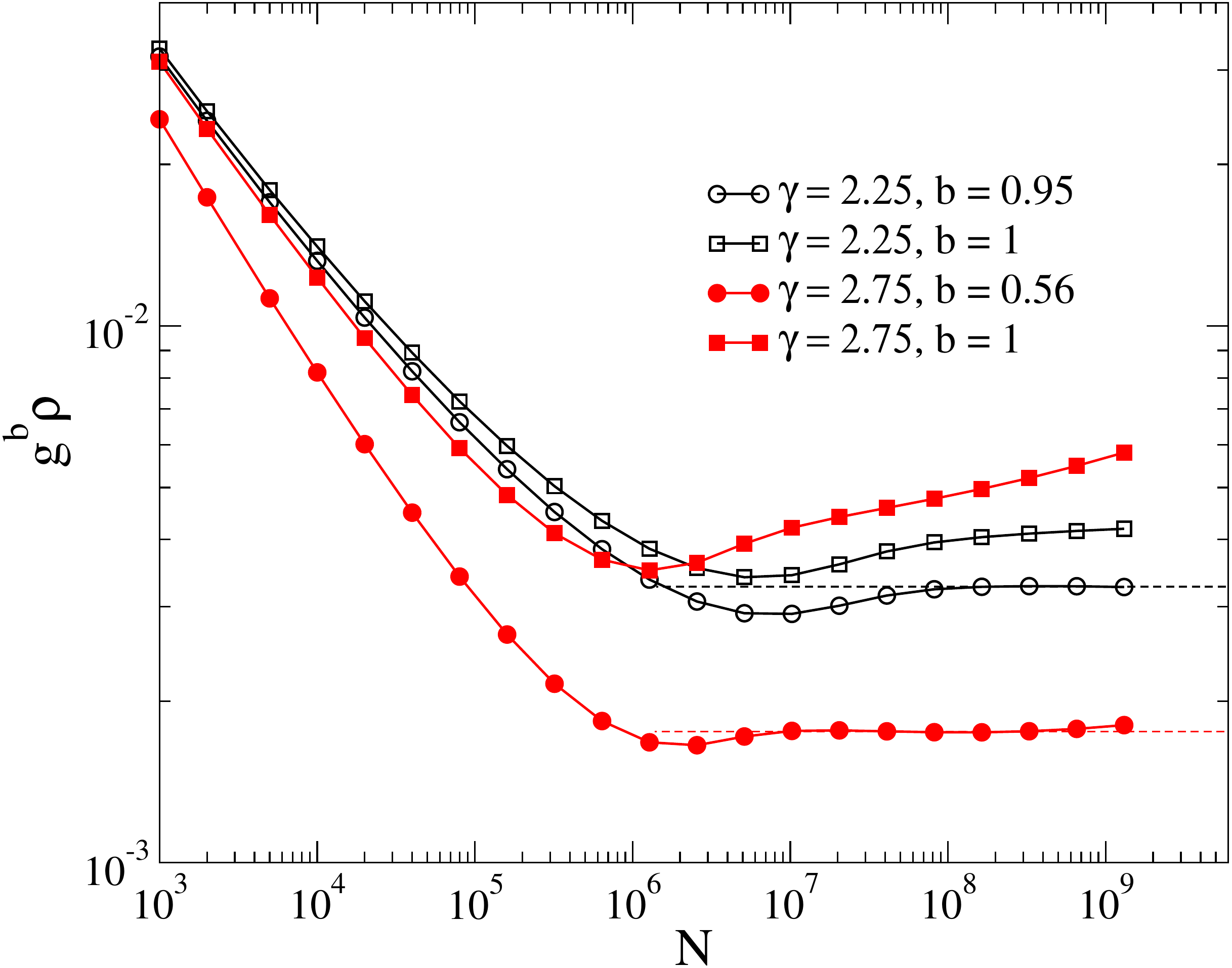}
  \caption{(Color on-line) Probing the anomalous FSS
    $\bar{\rho}\sim\Delta/g^b$ in the ME equation iterations of the
    supercritical CP for $\lambda=1.040$ (top) and 1.004 (bottom) and
    a cutoff exponent $\omega=2$. Dashed horizontal lines are guides
    to the eye.  }
 \label{fig:rho_gtob}
\end{figure}

The strong size dependence observed in Fig. \ref{fig:qs_p04900} for SF
networks ($\gamma=2.25$ and 2.75) but not for the homogeneous one
($\gamma = 3.25$) calls for an anomalous dependence of $\bar{\rho}$ on
$g$, as pointed out in the Langevin \cite{boguna09:_langev} and random
walk mapping \cite{PhysRevE.79.056115} approaches. Our numerical
approach allows a more detailed investigation of this issue.  Inspired
by the anomalous FSS prediction \cite{boguna09:_langev},
\begin{equation}
\bar{\rho}\sim \Delta/g,~~~ \Delta>\sqrt{g/N},
\label{eq:Deltag}
\end{equation}
we have performed a further test, analyzing the behavior of the QS
density, rescaled as $g^b\bar{\rho}$, as a function of $N$, as shown
in Fig. \ref{fig:rho_gtob}.  The theoretical exponent $b=1$, which
corresponds to the anomalous scaling in Eq.~(\ref{eq:Deltag}), does
not show up as a plateau in the plots of $g^b\bar{\rho}$ versus
$N$. Interestingly, plateaus are observed if an exponent $b < 1$ is
used instead.  For $\lambda=1.040$, plateaus are observed for $b =
0.88$ and $b=0.53$ for $\gamma=2.25$ and $\gamma=2.75$,
respectively. In an analysis performed closer to critical point, for
$\lambda=1.004$, the plateaus are observed with larger exponents $b =
0.95$ and $b=0.56$ for $\gamma=2.25$ and $\gamma=2.75$,
respectively. Notice that a scaling consistent with $b=1$ was obtained
for $\gamma=2.25$ but not for $\gamma=2.75$.  Actually, the anomalous
scaling \eqref{eq:Deltag}, derived from Eq.~\eqref{eq:anomalousFSS}, is
valid for $\sqrt{g/N}<\Delta\ll g\langle k \rangle/k_c$.  For
$\omega=2$, the right and left sides of this inequality scale as $g
\langle k \rangle /k_c \simeq c_\gamma N^{-(\gamma-2)/2}$ and
$\sqrt{g/N} \simeq \tilde{c}_\gamma N^{-(\gamma-1)/4}$, respectively, where
$c_\gamma = k_0^\gamma(\gamma-2)/(3-\gamma)$ and $\tilde{c}_\gamma^2 =
k_0^{\gamma-1}(\gamma-1)(\gamma-2)^2/(3-\gamma)$ are constants of the
same order and $c_\gamma>\tilde{c}_\gamma$.  If $\gamma$ is close to
$3$, the exponents involved in the lower and upper bounds of $\Delta$
are very close and we cannot make $\Delta$ sufficiently small to
fulfill the upper bound and still larger than the lower one, except
for numerically unaccessible large systems. Therefore, this anomalous
scaling can be clearly seen only for $\gamma$ close to 2. The scaling
forms with $b<1$ are thus metastable crossovers between the regimes
$\bar{\rho}\sim(gN)^{-1/2}$ and $\bar{\rho}\sim\Delta/g$, that can
last for decades, and could only be resolved by simulations in much
larger systems sizes than those considered in this work (up to
$N=10^9$ in the ME solutions).

\section{\label{sec:conclu} Concluding remarks}

The contact process (CP) on scale-free networks shows remarkably rich
features, even in the simple case of random annealed topologies. In
the present work we have explored the quasi-stationary (QS) properties
of this problem by combining QS numerical simulations and a master
equation approach applied to an approximate mapping to a one-step
process. The resulting master equation, apart from providing quite
accurate analytical approximations for the asymptotic shape of the QS activity
distribution at the critical point, can be very efficiently solved
numerically.  The QS distribution and the relevant QS quantities
(density of active sites and characteristic time) determined in this
way show an excellent agreement with direct QS numerical simulations
of the contact process, both at criticality and in the supercritical
regime.

The high accuracy of our data allows to identify strong corrections to
the scaling in the critical quantities that mask the correct finite
size scaling exponents obtained analytically by means of an exact
mean-field solution. Both critical density and characteristic time
show tenuous curvatures as functions of the network size $N$ due to
finite size corrections to scaling that may provide incorrect
exponents if a simple power law decay is assumed.  In annealed
networks, for which the critical point is exactly known, we can
determine the corrections to scaling analytically and thus recover the
theoretical exponents in the finite size analysis, including the
abrupt change when the network loses its SF property.  The analysis of
the supercritical region, on the other hand, hints that those finite
size corrections are also relevant for very large network sizes. Indeed, the
asymptotic scaling is observable only for extremely large values of
$N$, much larger than those possibly attainable with present day
computers.

{It is worth noticing that the QS analysis presented in this work
  is equivalent to the Langevin approach developed in
  Ref.~\cite{boguna09:_langev} in the limit of very low densities
  (i.e. in the critical region, see Appendix). The main difference
  is that the former starts from an approximation of the original
  dynamical processes [Eq.  (\ref{eq:ratesonestep})], while the latter
  represents an exact Langevin approach in the coarse-grained limit
  (density approximated by a continuous variable). The critical
  advantage of the present approach lies on the fact that our ME
  analysis allows to determine the critical properties in a very
  intuitive way, as well as to easily obtain highly accurate results,
  free from statistical errors, for all quantities of interest.}

Our work opens the path to a more detailed characterization of
absorbing phase transitions on scale-free networks in general, and 
the CP in particular. In the more realistic framework of
quenched networks, in which edges are frozen and do not change, this
goal may be however hindered by the interplay between corrections to
scaling and the usual lack of knowledge about the true position of the
critical point. In fact, the standard characterization of the QS state
by the usual procedure assuming a simple power law of the system size 
at the critical point may be affected by two sources
of errors: the analysis may be misleadingly done off the critical
point and/or be affected by important scaling corrections.  Further work in
this direction, following the proposed lines, might thus help to throw
light on the numerical assessment of the correct critical scaling of
absorbing phase transitions on heterogeneous networks.

\begin{acknowledgments}
  This work was partially supported by the Brazilian agencies CNPq and
  FAPEMIG.  S.C.F thanks the kind hospitality at the Departament de
  F\'{\i}sica i Enginyeria Nuclear/UPC.  R.P.-S.  acknowledges
  financial support from the Spanish MEC, under project
  FIS2010-21781-C02-01; the Junta de Andaluc\'{\i}a, under project
  No. P09-FQM4682; and additional support through ICREA Academia,
  funded by the Generalitat de Catalunya.
\end{acknowledgments}


\appendix

\section*{\label{sec:append} Appendix} 

The connection between the Langevin approach developed in
Ref. \cite{boguna09:_langev} and the one-step process
(\ref{eq:ratesonestep}) can be established by analyzing the respective
Fokker-Planck (FP) equations. The general form of a FP equation for a
stochastic variable $x$ is \cite{vankampen}
\begin{equation}
 \frac{\partial P(x,t)}{\partial t} = -\frac{\partial}{\partial x} 
 A(x)P(x,t) + \frac{\partial^2 }{\partial x^2}D(x)P(x,t)
\end{equation}
where $A(x)$ and $D(x)$ are the drift and the diffusion terms,
respectively.  The Langevin analysis in Ref. \cite{boguna09:_langev}
yielded $A(n)= n [\lambda-1-\lambda \Theta(n/N)]$ and $D(n)= 2\lambda
n \Lambda(n/N)$, where $\Theta$ is given by Eq. \eqref{eq:Theta} and
\begin{equation}
 \Lambda(\rho) = \sum_k \frac{k P(k)}{\langle k \rangle[1+\lambda k
   \rho/\langle k \rangle]^3}. 
\end{equation}
In turn, the drift and diffusion terms of the FP equation for an
arbitrary one-step process is given by $A(n)=W(n+1,n)-W(n-1,n)$ and
$D(n)=W(n+1,n)+W(n-1,n)$ \cite{vankampen}, respectively.
Equation~\eqref{eq:ratesonestep} results in exactly the same drift 
obtained in the Langevin approach, while the diffusion term takes
the form $D(n)=(1+\lambda)n+\lambda\Theta(n/N)$. At low densities, one
can expand $\Theta$ and $\Lambda$ to leading order and obtain
$D(n)\simeq 2n+\mathcal{O}(\Theta)$ in both cases. Therefore, Langevin
and ME approaches are equivalent in the low density limit.


\begin{thebibliography}{32}
\expandafter\ifx\csname natexlab\endcsname\relax\def\natexlab#1{#1}\fi
\expandafter\ifx\csname bibnamefont\endcsname\relax
  \def\bibnamefont#1{#1}\fi
\expandafter\ifx\csname bibfnamefont\endcsname\relax
  \def\bibfnamefont#1{#1}\fi
\expandafter\ifx\csname citenamefont\endcsname\relax
  \def\citenamefont#1{#1}\fi
\expandafter\ifx\csname url\endcsname\relax
  \def\url#1{\texttt{#1}}\fi
\expandafter\ifx\csname urlprefix\endcsname\relax\def\urlprefix{URL }\fi
\providecommand{\bibinfo}[2]{#2}
\providecommand{\eprint}[2][]{\url{#2}}

\bibitem[{\citenamefont{Albert and Barab\'asi}(2002)}]{barabasi02}
\bibinfo{author}{\bibfnamefont{R.}~\bibnamefont{Albert}} \bibnamefont{and}
  \bibinfo{author}{\bibfnamefont{A.-L.} \bibnamefont{Barab\'asi}},
  \bibinfo{journal}{Rev. Mod. Phys.} \textbf{\bibinfo{volume}{74}},
  \bibinfo{pages}{47} (\bibinfo{year}{2002}).

\bibitem[{\citenamefont{Dorogovtsev and Mendes}(2003)}]{mendesbook}
\bibinfo{author}{\bibfnamefont{S.~N.} \bibnamefont{Dorogovtsev}}
  \bibnamefont{and} \bibinfo{author}{\bibfnamefont{J.~F.~F.}
  \bibnamefont{Mendes}}, \emph{\bibinfo{title}{Evolution of networks: From
  biological nets to the {I}nternet and {WWW}}} (\bibinfo{publisher}{Oxford
  University Press}, \bibinfo{address}{Oxford}, \bibinfo{year}{2003}).

\bibitem[{\citenamefont{Newman}(2003)}]{newman2003saf}
\bibinfo{author}{\bibfnamefont{M.}~\bibnamefont{Newman}},
  \bibinfo{journal}{SIAM Review} \textbf{\bibinfo{volume}{45}},
  \bibinfo{pages}{167} (\bibinfo{year}{2003}).

\bibitem[{\citenamefont{Barrat et~al.}(2008)\citenamefont{Barrat,
  Barth\'{e}lemy, and Vespignani}}]{barratbook}
\bibinfo{author}{\bibfnamefont{A.}~\bibnamefont{Barrat}},
  \bibinfo{author}{\bibfnamefont{M.}~\bibnamefont{Barth\'{e}lemy}},
  \bibnamefont{and}
  \bibinfo{author}{\bibfnamefont{A.}~\bibnamefont{Vespignani}},
  \emph{\bibinfo{title}{Dynamical Processes on Complex Networks}}
  (\bibinfo{publisher}{Cambridge University Press},
  \bibinfo{address}{Cambridge}, \bibinfo{year}{2008}).

\bibitem[{\citenamefont{Dorogovtsev et~al.}(2008)\citenamefont{Dorogovtsev,
  Goltsev, and Mendes}}]{dorogovtsev07:_critic_phenom}
\bibinfo{author}{\bibfnamefont{S.~N.} \bibnamefont{Dorogovtsev}},
  \bibinfo{author}{\bibfnamefont{A.~V.} \bibnamefont{Goltsev}},
  \bibnamefont{and} \bibinfo{author}{\bibfnamefont{J.~F.~F.}
  \bibnamefont{Mendes}}, \bibinfo{journal}{Rev. Mod. Phys.}
  \textbf{\bibinfo{volume}{80}}, \bibinfo{pages}{1275} (\bibinfo{year}{2008}).

\bibitem[{\citenamefont{Anderson and May}(1992)}]{anderson92}
\bibinfo{author}{\bibfnamefont{R.~M.} \bibnamefont{Anderson}} \bibnamefont{and}
  \bibinfo{author}{\bibfnamefont{R.~M.} \bibnamefont{May}},
  \emph{\bibinfo{title}{Infectious diseases in humans}}
  (\bibinfo{publisher}{Oxford University Press}, \bibinfo{address}{Oxford},
  \bibinfo{year}{1992}).

\bibitem[{\citenamefont{Pastor-Satorras and Vespignani}(2004)}]{romuvespibook}
\bibinfo{author}{\bibfnamefont{R.}~\bibnamefont{Pastor-Satorras}}
  \bibnamefont{and}
  \bibinfo{author}{\bibfnamefont{A.}~\bibnamefont{Vespignani}},
  \emph{\bibinfo{title}{Evolution and structure of the Internet: A statistical
  physics approach}} (\bibinfo{publisher}{Cambridge University Press},
  \bibinfo{address}{Cambridge}, \bibinfo{year}{2004}).

\bibitem[{\citenamefont{Barrat et~al.}(2004)\citenamefont{Barrat,
  Barth\'{e}lemy, Pastor-Satorras, and Vespignani}}]{Barrat:2004b}
\bibinfo{author}{\bibfnamefont{A.}~\bibnamefont{Barrat}},
  \bibinfo{author}{\bibfnamefont{M.}~\bibnamefont{Barth\'{e}lemy}},
  \bibinfo{author}{\bibfnamefont{R.}~\bibnamefont{Pastor-Satorras}},
  \bibnamefont{and}
  \bibinfo{author}{\bibfnamefont{A.}~\bibnamefont{Vespignani}},
  \bibinfo{journal}{Proc. Natl. Acad. Sci. USA} \textbf{\bibinfo{volume}{101}},
  \bibinfo{pages}{3747} (\bibinfo{year}{2004}).

\bibitem[{\citenamefont{{Bogu\~{n}\'{a}} and Pastor-Satorras}(2002)}]{marian1}
\bibinfo{author}{\bibfnamefont{M.}~\bibnamefont{{Bogu\~{n}\'{a}}}}
  \bibnamefont{and}
  \bibinfo{author}{\bibfnamefont{R.}~\bibnamefont{Pastor-Satorras}},
  \bibinfo{journal}{Phys. Rev. E} \textbf{\bibinfo{volume}{66}},
  \bibinfo{pages}{047104} (\bibinfo{year}{2002}).

\bibitem[{\citenamefont{{Bogu\~{n}\'{a}}
  et~al.}(2009)\citenamefont{{Bogu\~{n}\'{a}}, Castellano, and
  Pastor-Satorras}}]{boguna09:_langev}
\bibinfo{author}{\bibfnamefont{M.}~\bibnamefont{{Bogu\~{n}\'{a}}}},
  \bibinfo{author}{\bibfnamefont{C.}~\bibnamefont{Castellano}},
  \bibnamefont{and}
  \bibinfo{author}{\bibfnamefont{R.}~\bibnamefont{Pastor-Satorras}},
  \bibinfo{journal}{Phys. Rev. E} \textbf{\bibinfo{volume}{79}},
  \bibinfo{pages}{036110} (\bibinfo{year}{2009}).

\bibitem[{\citenamefont{Marro and Dickman}(1999)}]{marro1999npt}
\bibinfo{author}{\bibfnamefont{J.}~\bibnamefont{Marro}} \bibnamefont{and}
  \bibinfo{author}{\bibfnamefont{R.}~\bibnamefont{Dickman}},
  \emph{\bibinfo{title}{{Nonequilibrium Phase Transitions in Lattice Models}}}
  (\bibinfo{publisher}{Cambridge University Press},
  \bibinfo{address}{Cambridge}, \bibinfo{year}{1999}).

\bibitem[{\citenamefont{Henkel et~al.}(2008)\citenamefont{Henkel, Hinrichsen,
  and L\"ubeck}}]{Henkel}
\bibinfo{author}{\bibfnamefont{M.}~\bibnamefont{Henkel}},
  \bibinfo{author}{\bibfnamefont{H.}~\bibnamefont{Hinrichsen}},
  \bibnamefont{and} \bibinfo{author}{\bibfnamefont{S.}~\bibnamefont{L\"ubeck}},
  \emph{\bibinfo{title}{Non-equilibrium phase transition: Absorbing Phase
  Transitions}} (\bibinfo{publisher}{Springer Verlag},
  \bibinfo{address}{Netherlands}, \bibinfo{year}{2008}).

\bibitem[{\citenamefont{Pastor-Satorras and Vespignani}(2001)}]{pv01a}
\bibinfo{author}{\bibfnamefont{R.}~\bibnamefont{Pastor-Satorras}}
  \bibnamefont{and}
  \bibinfo{author}{\bibfnamefont{A.}~\bibnamefont{Vespignani}},
  \bibinfo{journal}{Phys. Rev. Lett.} \textbf{\bibinfo{volume}{86}},
  \bibinfo{pages}{3200} (\bibinfo{year}{2001}).

\bibitem[{\citenamefont{Castellano and
  Pastor-Satorras}(2010)}]{PhysRevLett.105.218701}
\bibinfo{author}{\bibfnamefont{C.}~\bibnamefont{Castellano}} \bibnamefont{and}
  \bibinfo{author}{\bibfnamefont{R.}~\bibnamefont{Pastor-Satorras}},
  \bibinfo{journal}{Phys. Rev. Lett.} \textbf{\bibinfo{volume}{105}},
  \bibinfo{pages}{218701} (\bibinfo{year}{2010}).

\bibitem[{\citenamefont{Callaway et~al.}(2000)\citenamefont{Callaway, Newman,
  Strogatz, and Watts}}]{PhysRevLett.85.5468}
\bibinfo{author}{\bibfnamefont{D.~S.} \bibnamefont{Callaway}},
  \bibinfo{author}{\bibfnamefont{M.~E.~J.} \bibnamefont{Newman}},
  \bibinfo{author}{\bibfnamefont{S.~H.} \bibnamefont{Strogatz}},
  \bibnamefont{and} \bibinfo{author}{\bibfnamefont{D.~J.} \bibnamefont{Watts}},
  \bibinfo{journal}{Phys. Rev. Lett.} \textbf{\bibinfo{volume}{85}},
  \bibinfo{pages}{5468} (\bibinfo{year}{2000}).

\bibitem[{\citenamefont{Cohen et~al.}(2000)\citenamefont{Cohen, Erez,
  {ben-Avraham}, and Havlin}}]{PhysRevLett.85.4626}
\bibinfo{author}{\bibfnamefont{R.}~\bibnamefont{Cohen}},
  \bibinfo{author}{\bibfnamefont{K.}~\bibnamefont{Erez}},
  \bibinfo{author}{\bibfnamefont{D.}~\bibnamefont{{ben-Avraham}}},
  \bibnamefont{and} \bibinfo{author}{\bibfnamefont{S.}~\bibnamefont{Havlin}},
  \bibinfo{journal}{Phys. Rev. Lett.} \textbf{\bibinfo{volume}{85}},
  \bibinfo{pages}{4626} (\bibinfo{year}{2000}).

\bibitem[{\citenamefont{Goh et~al.}(2003)\citenamefont{Goh, Lee, Kahng, and
  Kim}}]{PhysRevLett.91.148701}
\bibinfo{author}{\bibfnamefont{K.-I.} \bibnamefont{Goh}},
  \bibinfo{author}{\bibfnamefont{D.-S.} \bibnamefont{Lee}},
  \bibinfo{author}{\bibfnamefont{B.}~\bibnamefont{Kahng}}, \bibnamefont{and}
  \bibinfo{author}{\bibfnamefont{D.}~\bibnamefont{Kim}},
  \bibinfo{journal}{Phys. Rev. Lett.} \textbf{\bibinfo{volume}{91}},
  \bibinfo{pages}{148701} (\bibinfo{year}{2003}).

\bibitem[{\citenamefont{Harris}(1974)}]{harris74}
\bibinfo{author}{\bibfnamefont{T.~E.} \bibnamefont{Harris}},
  \bibinfo{journal}{Ann. Prob.} \textbf{\bibinfo{volume}{2}},
  \bibinfo{pages}{969} (\bibinfo{year}{1974}).

\bibitem[{\citenamefont{Castellano and
  Pastor-Satorras}(2006)}]{Castellano:2006}
\bibinfo{author}{\bibfnamefont{C.}~\bibnamefont{Castellano}} \bibnamefont{and}
  \bibinfo{author}{\bibfnamefont{R.}~\bibnamefont{Pastor-Satorras}},
  \bibinfo{journal}{Phys. Rev. Lett.} \textbf{\bibinfo{volume}{96}},
  \bibinfo{pages}{038701} (\bibinfo{year}{2006}).

\bibitem[{\citenamefont{Hong et~al.}(2007)\citenamefont{Hong, Ha, and
  Park}}]{Hong:2007}
\bibinfo{author}{\bibfnamefont{H.}~\bibnamefont{Hong}},
  \bibinfo{author}{\bibfnamefont{M.}~\bibnamefont{Ha}}, \bibnamefont{and}
  \bibinfo{author}{\bibfnamefont{H.}~\bibnamefont{Park}},
  \bibinfo{journal}{Phys. Rev. Lett.} \textbf{\bibinfo{volume}{98}},
  \bibinfo{pages}{258701} (\bibinfo{year}{2007}).

\bibitem[{\citenamefont{Castellano and
  Pastor-Satorras}(2008)}]{Castellano:2008}
\bibinfo{author}{\bibfnamefont{C.}~\bibnamefont{Castellano}} \bibnamefont{and}
  \bibinfo{author}{\bibfnamefont{R.}~\bibnamefont{Pastor-Satorras}},
  \bibinfo{journal}{Phys. Rev. Lett.} \textbf{\bibinfo{volume}{100}},
  \bibinfo{pages}{148701} (\bibinfo{year}{2008}).

\bibitem[{\citenamefont{Noh and Park}(2009)}]{PhysRevE.79.056115}
\bibinfo{author}{\bibfnamefont{J.~D.} \bibnamefont{Noh}} \bibnamefont{and}
  \bibinfo{author}{\bibfnamefont{H.}~\bibnamefont{Park}},
  \bibinfo{journal}{Phys. Rev. E} \textbf{\bibinfo{volume}{79}},
  \bibinfo{pages}{056115} (\bibinfo{year}{2009}).

\bibitem[{\citenamefont{Cardy}(1988)}]{cardy88}
\bibinfo{editor}{\bibfnamefont{J.~L.} \bibnamefont{Cardy}}, ed.,
  \emph{\bibinfo{title}{Finite Size Scaling}}, vol.~\bibinfo{volume}{2}
  (\bibinfo{publisher}{North Holland}, \bibinfo{address}{Amsterdam},
  \bibinfo{year}{1988}).

\bibitem[{\citenamefont{de~Oliveira and Dickman}(2005)}]{PhysRevE.71.016129}
\bibinfo{author}{\bibfnamefont{M.~M.} \bibnamefont{de~Oliveira}}
  \bibnamefont{and} \bibinfo{author}{\bibfnamefont{R.}~\bibnamefont{Dickman}},
  \bibinfo{journal}{Phys. Rev. E} \textbf{\bibinfo{volume}{71}},
  \bibinfo{pages}{016129} (\bibinfo{year}{2005}).

\bibitem[{\citenamefont{Dickman and Vidigal}(2002)}]{DickmanJPA}
\bibinfo{author}{\bibfnamefont{R.}~\bibnamefont{Dickman}} \bibnamefont{and}
  \bibinfo{author}{\bibfnamefont{R.}~\bibnamefont{Vidigal}},
  \bibinfo{journal}{J. Phys. A: Math. Gen.} \textbf{\bibinfo{volume}{35}},
  \bibinfo{pages}{1147} (\bibinfo{year}{2002}).

\bibitem[{\citenamefont{Watts and Strogatz}(1998)}]{watts98}
\bibinfo{author}{\bibfnamefont{D.~J.} \bibnamefont{Watts}} \bibnamefont{and}
  \bibinfo{author}{\bibfnamefont{S.~H.} \bibnamefont{Strogatz}},
  \bibinfo{journal}{Nature} \textbf{\bibinfo{volume}{393}},
  \bibinfo{pages}{440} (\bibinfo{year}{1998}).

\bibitem[{\citenamefont{Dickman}(2006)}]{PhysRevE.73.036131}
\bibinfo{author}{\bibfnamefont{R.}~\bibnamefont{Dickman}},
  \bibinfo{journal}{Phys. Rev. E} \textbf{\bibinfo{volume}{73}},
  \bibinfo{pages}{036131} (\bibinfo{year}{2006}).

\bibitem[{\citenamefont{de~Oliveira et~al.}(2008)\citenamefont{de~Oliveira,
  Alves, Ferreira, and Dickman}}]{PhysRevE.78.031133}
\bibinfo{author}{\bibfnamefont{M.~M.} \bibnamefont{de~Oliveira}},
  \bibinfo{author}{\bibfnamefont{S.~G.} \bibnamefont{Alves}},
  \bibinfo{author}{\bibfnamefont{S.~C.} \bibnamefont{Ferreira}},
  \bibnamefont{and} \bibinfo{author}{\bibfnamefont{R.}~\bibnamefont{Dickman}},
  \bibinfo{journal}{Phys. Rev. E} \textbf{\bibinfo{volume}{78}},
  \bibinfo{pages}{031133} (\bibinfo{year}{2008}).

\bibitem[{\citenamefont{{Bogu\~{n}\'{a}}
  et~al.}(2004)\citenamefont{{Bogu\~{n}\'{a}}, Pastor-Satorras, and
  Vespignani}}]{mariancutofss}
\bibinfo{author}{\bibfnamefont{M.}~\bibnamefont{{Bogu\~{n}\'{a}}}},
  \bibinfo{author}{\bibfnamefont{R.}~\bibnamefont{Pastor-Satorras}},
  \bibnamefont{and}
  \bibinfo{author}{\bibfnamefont{A.}~\bibnamefont{Vespignani}},
  \bibinfo{journal}{Euro. Phys. J. B} \textbf{\bibinfo{volume}{38}},
  \bibinfo{pages}{205} (\bibinfo{year}{2004}).

\bibitem[{\citenamefont{{van Kampen}}(1981)}]{vankampen}
\bibinfo{author}{\bibfnamefont{N.~G.} \bibnamefont{{van Kampen}}},
  \emph{\bibinfo{title}{Stochastic processes in chemistry and physics}}
  (\bibinfo{publisher}{North Holland}, \bibinfo{address}{Amsterdam},
  \bibinfo{year}{1981}).

\bibitem[{\citenamefont{Castellano and
  Pastor-Satorras}(2007)}]{PhysRevLett.98.029802}
\bibinfo{author}{\bibfnamefont{C.}~\bibnamefont{Castellano}} \bibnamefont{and}
  \bibinfo{author}{\bibfnamefont{R.}~\bibnamefont{Pastor-Satorras}},
  \bibinfo{journal}{Phys. Rev. Lett.} \textbf{\bibinfo{volume}{98}},
  \bibinfo{pages}{029802} (\bibinfo{year}{2007}).

\bibitem[{\citenamefont{Sander et~al.}(2009)\citenamefont{Sander, de~Oliveira,
  and Ferreira}}]{Sander2009}
\bibinfo{author}{\bibfnamefont{R.~S.} \bibnamefont{Sander}},
  \bibinfo{author}{\bibfnamefont{M.~M.} \bibnamefont{de~Oliveira}},
  \bibnamefont{and} \bibinfo{author}{\bibfnamefont{S.~C.}
  \bibnamefont{Ferreira}}, \bibinfo{journal}{J. Stat. Mech. Theor. Exp.}
  \textbf{\bibinfo{volume}{2009}}, \bibinfo{pages}{P08011}
  (\bibinfo{year}{2009}).

\end{thebibliography}

\end{document}